\documentclass[
 reprint,
 amsmath,amssymb,
 aps,
]{revtex4-1}

\usepackage{subcaption}
\usepackage{dcolumn}
\usepackage{bm}
\usepackage{amsmath}
\usepackage{physics}
\usepackage{multirow}
\usepackage{graphicx}

\begin{document}

\preprint{APS/123-QED}

\title{Analysis of Measurement-based Quantum Network Coding \\ over Repeater Networks under Noisy Conditions}

\author{Takaaki Matsuo}
\altaffiliation{Keio University Shonan Fujisawa Campus, 5322 Endo, Fujisawa, Kanagawa 252-0882, Japan}
\author{Takahiko Satoh}
\altaffiliation{Keio University Shonan Fujisawa Campus, 5322 Endo, Fujisawa, Kanagawa 252-0882, Japan}
\author{Shota Nagayama}
\altaffiliation{Keio University Shonan Fujisawa Campus, 5322 Endo, Fujisawa, Kanagawa 252-0882, Japan}
\altaffiliation{Currently at R4D, Mercari, Inc.}
\author{Rodney Van Meter}
\altaffiliation{Keio University Shonan Fujisawa Campus, 5322 Endo, Fujisawa, Kanagawa 252-0882, Japan}

\date{\today}

\begin{abstract}
Quantum network coding is an effective solution for alleviating bottlenecks in quantum networks. We introduce a measurement-based quantum network coding scheme for quantum repeater networks (MQNC), and analyze its behavior based on results acquired from Monte-Carlo simulation that includes various error sources over a butterfly network.
By exploiting measurement-based quantum computing, operation on qubits for completing network coding proceeds in parallel.
We show that such an approach offers advantages over other schemes in terms of the quantum circuit depth, and therefore improves the communication fidelity without disturbing the aggregate throughput.
The circuit depth of our protocol has been reduced by 56.5\% compared to the quantum network coding scheme (QNC)
introduced in 2012 by Satoh, \emph{et al.} For MQNC, we have found that the resulting entangled pairs' joint fidelity drops below 50\% when the accuracy of local operations is lower than 98.9\%, assuming that all initial Bell pairs across quantum repeaters have a fixed fidelity of 98\%. Overall, MQNC showed substantially higher error tolerance compared to QNC and slightly better than  buffer space multiplexing using step-by-step entanglement swapping, but not quite as strong as simultaneous entanglement swapping operations.
\end{abstract}

\maketitle

\section{\label{sec:intro}INTRODUCTION}
Quantum network coding is a promising technique used for improving the aggregate throughput of a quantum network.  Linear operations on data at nodes in the middle of the network allow efficient exchange of quantum information, alleviating bottlenecks caused by topological constraints.

\begin{figure}[htbp]
  \center
  \includegraphics[keepaspectratio,scale=0.45]{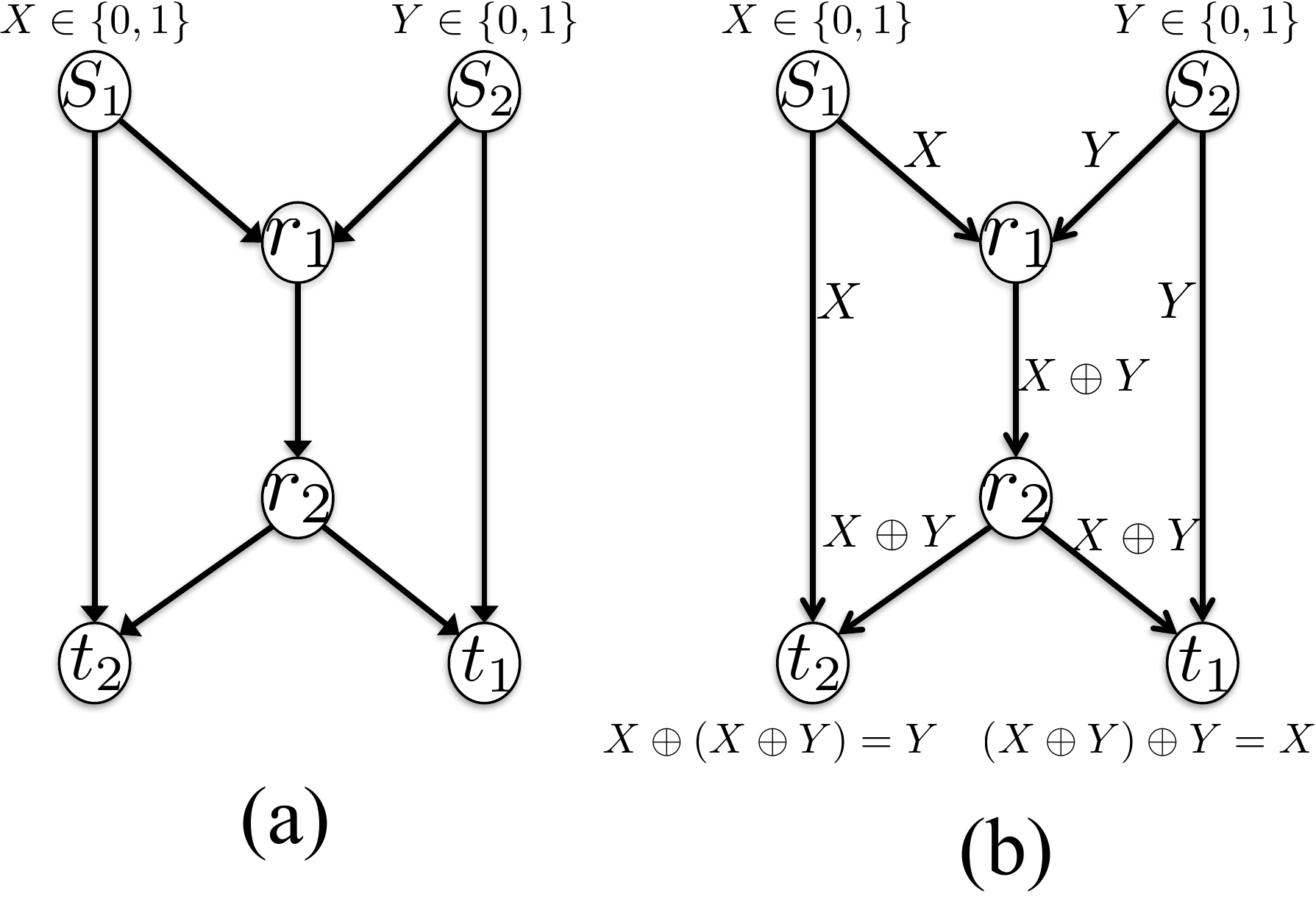}
  \caption{Fundamental network topology with bottleneck solvable via network coding. (a) The butterfly network with a bottleneck at link between intermediate resource nodes $r_{1}$ and $r_{2}$. Even with undirected channels, resource contention occurs somewhere with standard routing protocol. (b) Network coding performed to transmit two messages simultaneously. Messages are encoded at resource node $r_{1}$ and decoded at target nodes $t_{1}$ and $t_{2}$.}
  \label{The_Butterfly}
\end{figure}

Network coding, proposed by Ahlswede, Cai, Li, and Yeung, is a classical communication technique used for alleviating bottlenecks in a classical network \cite{Ahlswede2000}. Unlike the standard switch used for routing, network coding requires additional node functionality, using an encoder and decoder dedicated for linearly combining more than one message and for reversibly constructing the original message, which can be used to improve the network throughput for certain traffic patterns. The simplest example of network coding can be explained over a butterfly network as illustrated in Fig. \ref{The_Butterfly}(a).

In this example, there are two source nodes $S_{1}$  and $S_{2}$ with the goal of delivering messages $X$ and $Y$ to their target nodes $t_{1}$ and $t_{2}$ respectively. Here, each message is assumed to be 1 bit of data, and all directed channels have a limited capacity of 1 bit per unit time. With a general routing protocol, no matter what path is chosen for each connection, the two paths must overlap somewhere, resulting in contention for access to one link.
Therefore, the link between the intermediate resource nodes $r_{1}$ and $r_{2}$ becomes a bottleneck.
One possible solution for such a problem may be the use of time division multiplexing, which uses two cycles to complete the message transmissions.

\begin{figure*}[htbp]
  \center
  \includegraphics[keepaspectratio,scale=0.4]{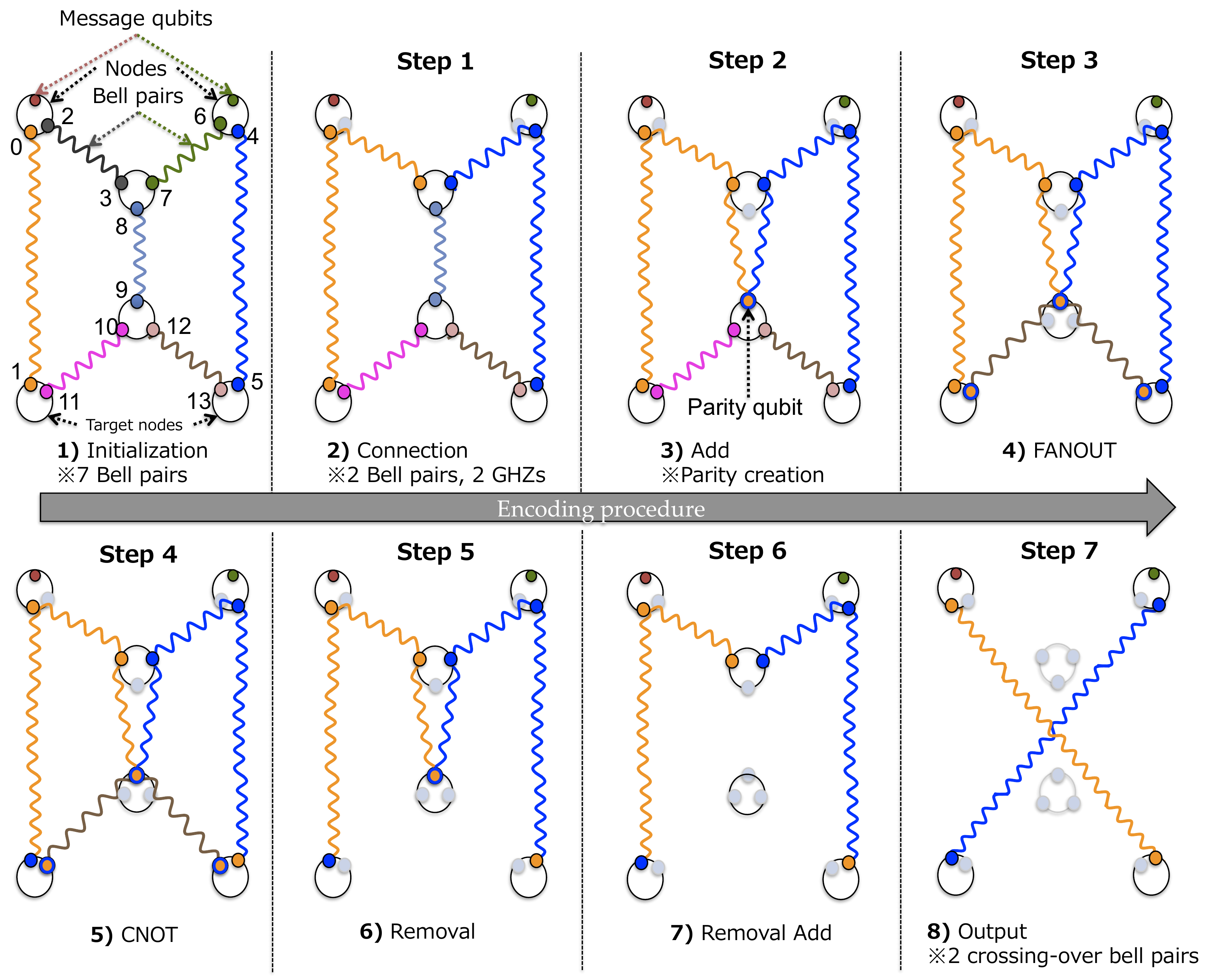}
  \caption{Step-by-step encoding procedure of QNC. This scheme is performs network coding on quantum channels across repeaters in a similar way as the classical network coding algorithm. The qubit at the bottleneck will be manipulated to be the parity of 2 GHZ states to the left and right.}
  \label{qnc}
\end{figure*}

On the other hand, network coding is capable of completing both transmissions within one cycle by linearly combining the incoming messages and transmitting them as a single message (see Fig. \ref{The_Butterfly}(b)). Source node $S_{1}$, which wants to send message $X$ to $t_{1}$, sends its message towards the target node $t_{2}$ and to the resource node $r_{1}$. Similarly, source node $S_{2}$ forwards its message towards the target node $t_{1}$ and the resource node $r_{1}$. The resource node $r_{1}$ then processes the incoming messages, using an XOR operation to linearly combine the messages, and forwards the encoded message to both target nodes via $r_{2}$. At the end, each node can reconstruct their desired message by decoding the linearly combined message, using XOR operation between the other message directly sent from the source node to the target node. Hence, network coding can achieve a throughput of two messages per cycle.

To tackle similar issues that occur in quantum networks, a number of quantum network coding techniques have been proposed.
Many protocols do not address noisy operations or decoherence \cite{2006quant.ph..1088H,Kobayashi2009, Kobayashi2010, Kobayashi2011, Beaudrap2014}, and assume the ability to transmit qubits perfectly along channels from node to node with or without classical support.

Some assume lossy and noisy channels with imperfect gate operations.
One of the protocols for quantum network coding is motivated by Bell pair-based quantum repeater networks \cite{Satoh2012}. The error tolerance of the proposed protocol in \cite{Satoh2012} is analyzed over various errors using Monte-Carlo simulation with discussions on the advantages and disadvantages of network coding relative to the standard routing strategy based on entanglement swapping \cite{Satoh2016}.
Another proposal focuses on graph state networks composed of quantum routers with the ability to perform basic measurement-based quantum computations, which also employs network coding \cite{Epping2016}. Epping \emph{et al.} consider depolarizing channels and analyzes the error correction capabilities of quantum network coding in the context of stabilizer codes and stabilizer error correction codes and discusses the robustness of network coding.

In this paper, we simplified the protocol in \cite{Satoh2012} by adopting measurement-based quantum computing in order to process operations in parallel, but without loss of capability. We study the creation of the necessary cluster state building on repeater-produced two-qubit cluster states, and find that this approach consumes fewer resources and results in higher fidelity than the Bell pair-based approach.

\section{RELATED WORK}

\subsection{Quantum Network Coding}

Hayashi \emph{et al.} first introduced quantum network coding in 2006 \cite{2006quant.ph..1088H}. They focused on the theoretical approach of quantum network coding without classical communication support, and showed that the communication fidelity is upper bounded by $F_{output}<0.983$, when simultaneously transmitting arbitrary quantum states over a butterfly network via quantum network coding.
In 2007, Leung \emph{et al.} generalized the impossibility of perfect quantum network coding to several network types beyond the butterfly network, and showed that perfect quantum network coding is impossible even with asymptotically perfect transmission \cite{Winter2007}.
For each topology, they have also assumed different kinds of supporting classical channels that include forward-assisted, back-assisted and two-way assisted channels, and concluded that the communication fidelity supported with the two-way assisted classical channels is no better than the case with the free classical back-assisted channels, and in all cases, there is an upper bound and a lower bound for the communication fidelity.
Later on, Kobayashi \emph{et al.} showed that perfect quantum network coding can be accomplished whenever free classical channels are available, for any graph shape that is solvable by classical network coding \cite{Kobayashi2009, Kobayashi2010, Kobayashi2011}.
Beaudrap and Roetteler showed that the classically assisted quantum network coding scheme in \cite{Kobayashi2009, Kobayashi2011} corresponds to measurement-based quantum computing \cite{Beaudrap2014}.
All of these results assumed that qubits are immune to noise.

\subsection{Quantum Repeater Network Coding (QNC)}
Quantum repeaters, introduced by Briegel \emph{et al.} in the late 1990s, are a promising technology for enabling multi-hop quantum communications and managing errors using entangled qubits distributed over long distances \cite{Briegel1998, Briegel1999}. Knill and Laflamme introduced an error correction based fault-tolerant quantum communication scheme in 1996 \cite{1996quant.ph..8012K}.

The ability to manipulate the quantum channels across repeaters also allows us to complete network coding without disturbing the message qubit until the very end of the protocol.
Such a method was proposed by Satoh \emph{et al.} in 2012, as a network coding protocol for noisy and lossy quantum repeater networks (QNC) \cite{Satoh2012}.
Performing complex gate operations directly on the message qubits degrades the qubit state, and performing purification on the message after a complex encoding may not be easy. Instead, QNC focuses on creating two end-to-end Bell pairs between the source-destination pairs by consuming the entangled resources shared across the repeaters with a goal of lowering the protocol complexity and thus improving the communication fidelity.

As shown in the QNC encoding procedure in Fig. \ref{qnc} and the corresponding circuit in Fig. \ref{circuits}(a), the network is assumed to have seven Bell pairs shaping a butterfly graph. Classical channels are assumed to be undirected and have unlimited capacity. With the given seven Bell pairs across six nodes, operations convert the given resources into two independent Bell pairs from source to target directly, which can be used to teleport the message qubit to the desired destination (refer to Appendix \ref{qnc_step} for details).

In 2016, Satoh \emph{\emph{et al.}} studied the behavior of QNC under noisy conditions using Monte-Carlo simulation and compared it with the standard routing technique using entanglement swapping \cite{Satoh2016}. Their paper concluded that the routing protocol tolerates about twice the local error rate of QNC.
Each operation in QNC is ordered in time, therefore, qubit dependencies worsen the quantum circuit depth.
Due to the high circuit complexity, local operation accuracies tend to have a larger impact on the output fidelity compared to pre-shared entangled resource fidelity.
Moreover, even with perfect local gates, the output fidelity drops below $F_{output} < 0.5$ when Bell pairs have fidelity $F_{input}<0.90$.
While the standard routing protocol offers higher communication fidelity, QNC reduces the required number of cycles, and therefore provides a benefit if network resources are limited or if higher communication speed is demanded.

\subsection{Measurement-based Quantum Computing (MBQC)}
Measurement-based Quantum Computing (MBQC) is an alternative universal computation method based on single qubit measurements on a cluster state, which was proposed by Raussendorf \emph{et al.} in 2003 \cite{2003PhRvA..68b2312R}.

A cluster state of $n$ vertices (qubits) can be defined by:
\begin{equation}
\ket{G} = \prod_{(a,b)\in E}\Lambda_{a,b}(Z)\ket{+}^{\otimes n}
\end{equation}

where E is the set of edges (entanglement) and $a,b$ are the corresponding vertices (qubits).

\begin{figure}[htbp]
  \center
  \includegraphics[keepaspectratio,scale=0.35]{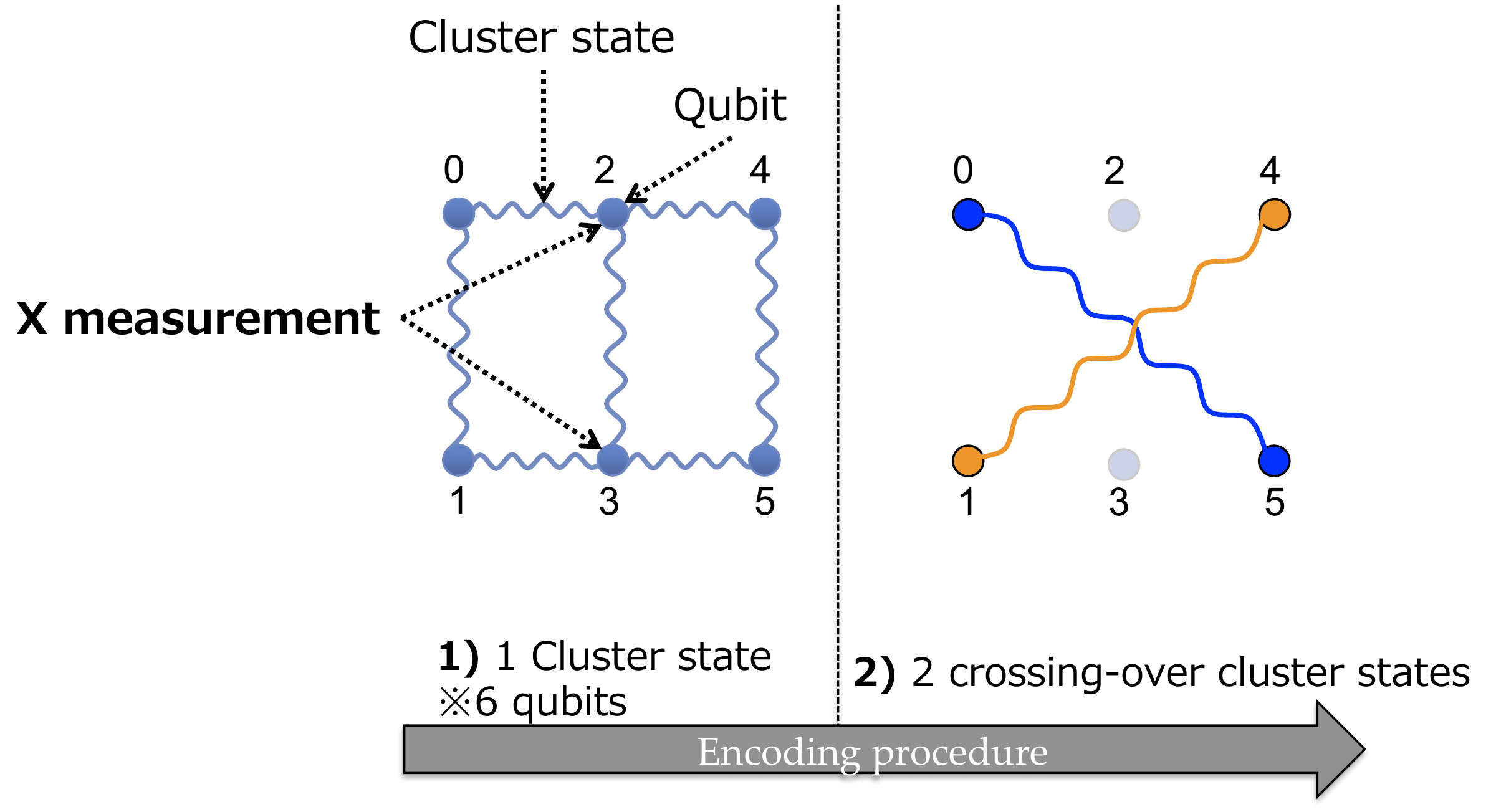}
  \caption{Visualized encoding procedure of network coding on a cluster state. This scheme may be used as a swap gate for MBQC.}
  \label{mbqc_qnc}
\end{figure}

Performing X-measurements on the bottleneck qubits of a butterfly cluster state will result in two cross-over two-qubit cluster states \cite{Epping2016}.
For an illustrated model of this scheme, see Fig. \ref{mbqc_qnc}.
Although this algorithm seems to be simpler than QNC, one should be reminded that creating a cluster state requires pairwise entanglement of all qubits, thus, it is not feasible to directly create a multi-qubit cluster state using qubits that are far apart. The physical system used for MBQC often is assumed to be a system area network.

\section{Measurement-based Quantum Repeater Network Coding (MQNC)}

\begin{figure*}[!hbt]
  \center
  \includegraphics[keepaspectratio,scale=0.4]{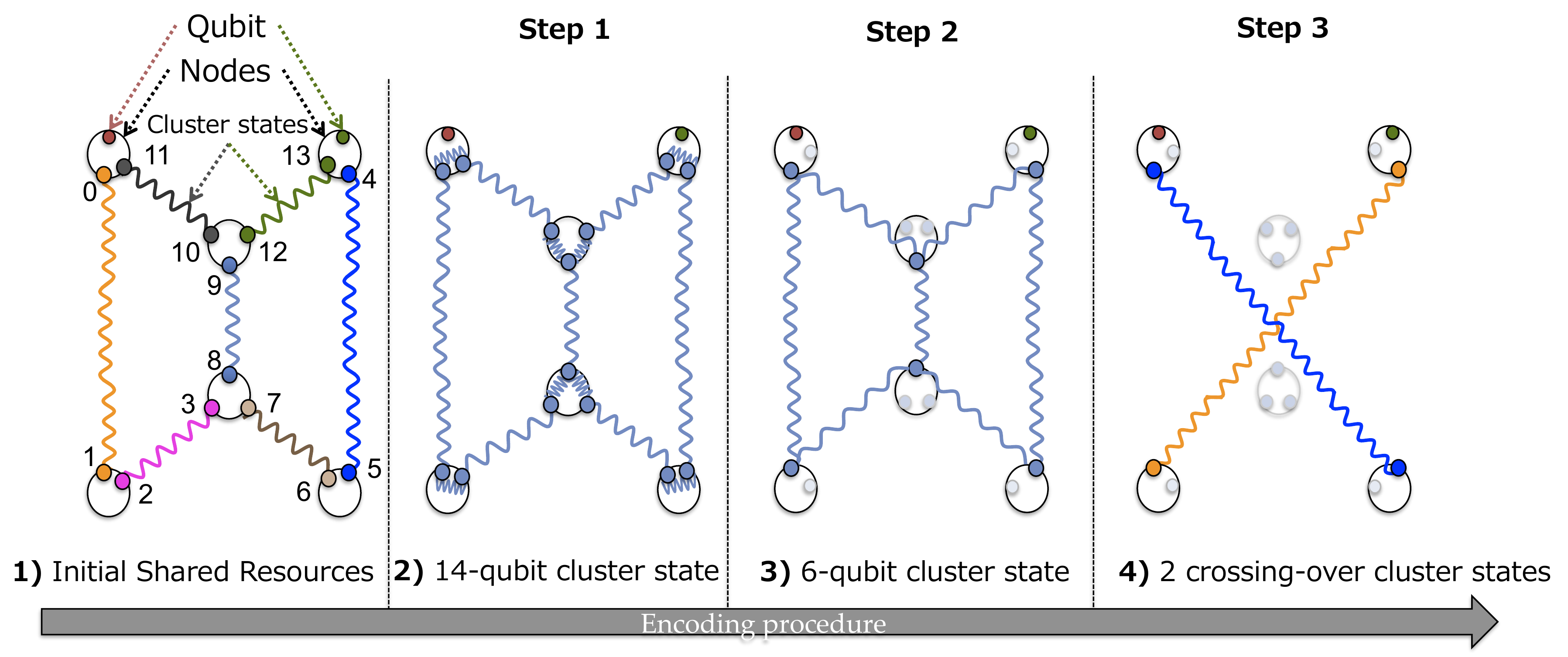}
  \caption{Step-by-step encoding procedure of MQNC. This scheme also manipulates quantum channels but without any parity creation. The topological transition via measurements on cluster states can accomplish the same goal as QNC in a simpler way.}
  \label{mqnc_encoding_full}
\end{figure*}

Recently introduced quantum network coding protocols are generally designed based on a classical algorithm using the CNOT operation, which is the quantum equivalent of XOR, and have high circuit complexities compared to the standard quantum routing protocol using entanglement swapping. Consequently, prior work on quantum network coding acquires higher aggregate network throughput but with a penalty on the communication fidelity due to complex operations.

The benefit of QNC comes from its applicability over quantum repeater networks and used for long distance communications. Nevertheless, the encoding procedure is still based on the classical counterpart which results in many qubit dependencies, lengthening the circuit depth, and therefore adversely affecting the communication fidelity. While the benefit of MBQC comes from the simplicity of implementation, the scheme for MBQC generally assumes a system area network.

We optimized the procedure of network coding for lossy and noisy repeater networks based on two technologies. First, parallelizing operations through the adaptation of MBQC. Second, combining conditional byproduct operations to suppress the influence of gate errors.
Our protocol, MQNC, takes advantage of both QNC and MBQC using local operations and classical communication (LOCC) and the entangled pairs created on repeater network links.
Unlike the other network coding schemes for quantum communication that directly encode qubits to combine messages, and the classical network coding based algorithms performed on quantum channels, MQNC focuses on generating two end-to-end paths, which partially takes the idea of QNC but without a single use of parity measurements. The basic idea of MQNC is to create a 6-qubit butterfly cluster state from the seven shared entangled pairs, and to treat the generated state as a resource state for network level MBQC, which allows us to topologically achieve the same goal as QNC. The developed protocol's pictorial model is shown in Fig. \ref{mqnc_encoding_full}.
Two-qubit entangled states across nodes are assumed to be ready, and therefore the link-timing architecture \cite{Codey2016, Liang2016} for entanglement distributions has not been taken into consideration here.

The encoding procedure for MQNC can be divided into three major steps (see Fig. \ref{mqnc_encoding_full}), with an assumption of accessible entangled resources across quantum repeaters:
\begin{equation}
  \ket{\Psi_{0}} = \ket{G_{0,1}}\ket{G_{2,3}}\ket{G_{4,5}}\ket{G_{6,7}}\ket{G_{8,9}}\ket{G_{10,11}}\ket{G_{12,13}}
\end{equation}

where $\ket{G_{0,1}}$ denotes a cluster state of qubit 0 and qubit 1.

The first step simply connects all local qubits via CZ operations in parallel, forming a single cluster state.

The second step is the creation of the butterfly-shaped 6-qubit cluster state using only LOCC and the resources prepared in the first step. Qubits are removed via $Y$ measurements and the neighboring qubits are directly connected up to the byproduct phase operations.

The last step completes the measurement-based quantum network coding by creating two cross-over independent cluster states out of the butterfly graph. Qubit 9 and qubit 8 at the bottleneck are measured with respect to the X-basis.

\begin{table}
\caption{Stabilizer set of qubit 0-5 after X-measurements in Step 3. ${t_{8}}$ and ${t_{9}}$ is the measurement outcome of qubit 8 and 9 respectively.}
\label{step3}
\begin{ruledtabular}
\begin{tabular}{c c c c c c c}
 &&&Qubits&&&\\ \hline \\
Stabilizer 1 & $X_{0}$ & & $-1^{t_{8}}$ & &  & $Z_{5}$ \\
Stabilizer 2 & & $X_{1}$ & & $-1^{t_{9}}$  & $Z_{4}$ &   \\
Stabilizer 3 &  & $Z_{1}$ & $-1^{t_{8}}$ & & $X_{4}$ &  \\
Stabilizer 4 & $Z_{0}$ & & & $-1^{t_{9}}$ &  & $X_{5}$   \\
\end{tabular}
\end{ruledtabular}
\end{table}

\begin{figure}[!hbt]
  \center
  \includegraphics[keepaspectratio,scale=0.35]{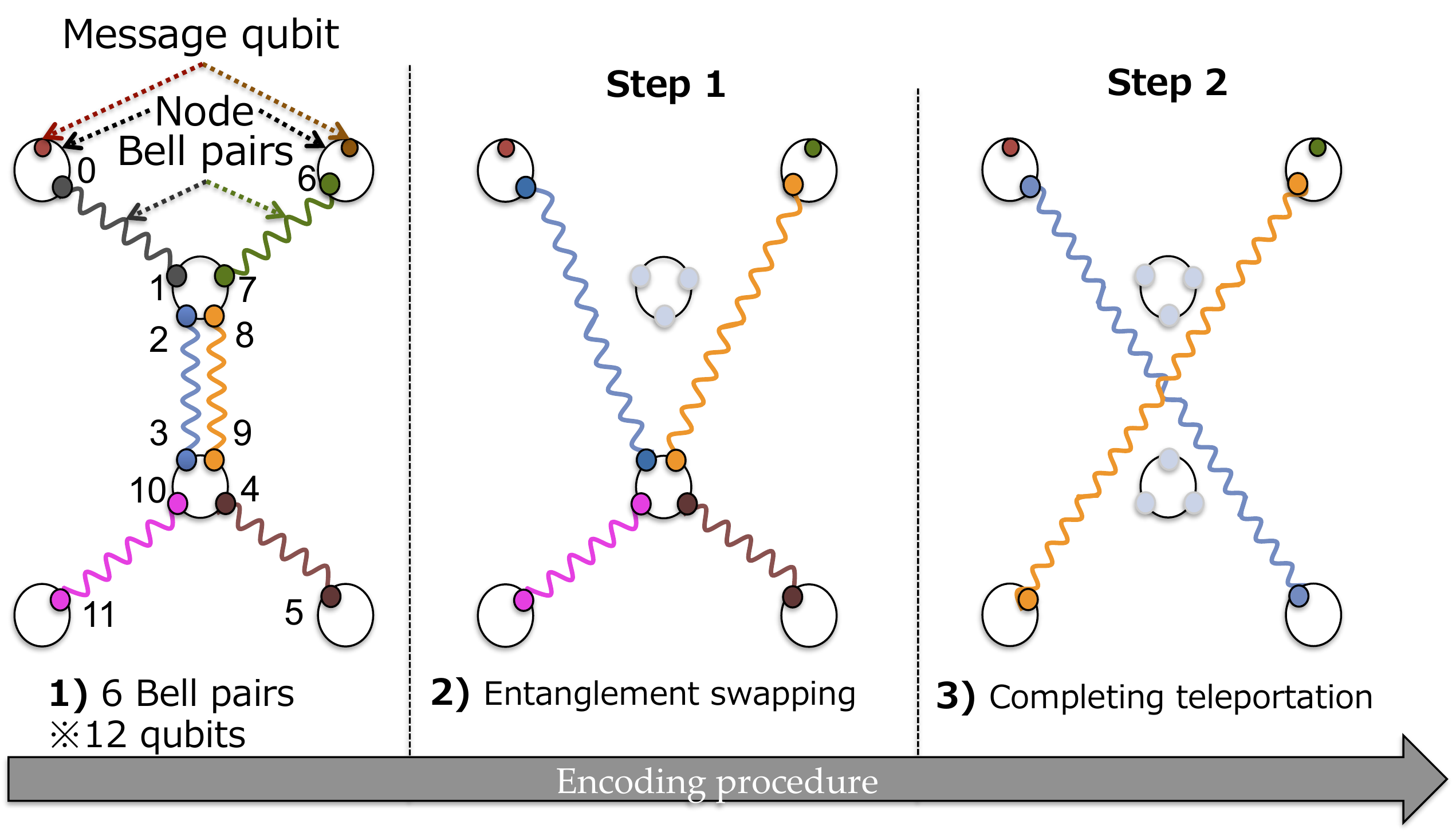}
  \caption{Buffer space multiplexing using entanglement swapping (ES). Each resource at the center link is assigned for completing different communications simultaneously. Step 1 can be omitted by performing all entanglement swapping operations simultaneously ($\mathrm{ES_{p}}$).}
  \label{buffer-space}
\end{figure}

As shown in Table \ref{step3}, in order to fix the phase to a desired state, either $Z$ gates or $X$ gates can be performed. When using $Z$ gates, two $Z$ gates are applied to qubit 0 and qubit 4 as a conditional byproduct of qubit 8 measurement.
Similarly, two $Z$ gates are applied to qubit 1 and qubit 5 as a conditional byproduct of measurement on qubit 9.
Alternatively, one can achieve the same goal by performing $X$ gates on qubit 0 and qubit 4 as a byproduct of measuring qubit 9, and on qubit 1 and qubit 5 as a byproduct of measuring qubit 8.

Using the $X$ byproduct operators is preferred, for the simple reason that they require slightly less classical communication.
The $X$ operators are applied one hop away from where the measurement occurs, whereas the $Z$ byproduct operators would be applied two hops away from the measurement operation.

Taking full advantage of the gate commutativity allows us to parallelize some encoding operations, which contributes to reducing the circuit depth.
All CZ gate operations can be applied in a parallel manner at the beginning of the protocol, and measuring qubits can be done afterwards. As a result, MQNC has achieved a 56.5\% reduction of circuit depth compared to QNC.

\begin{figure*}[!hbt]
  \center
  \includegraphics[keepaspectratio,scale=0.35]{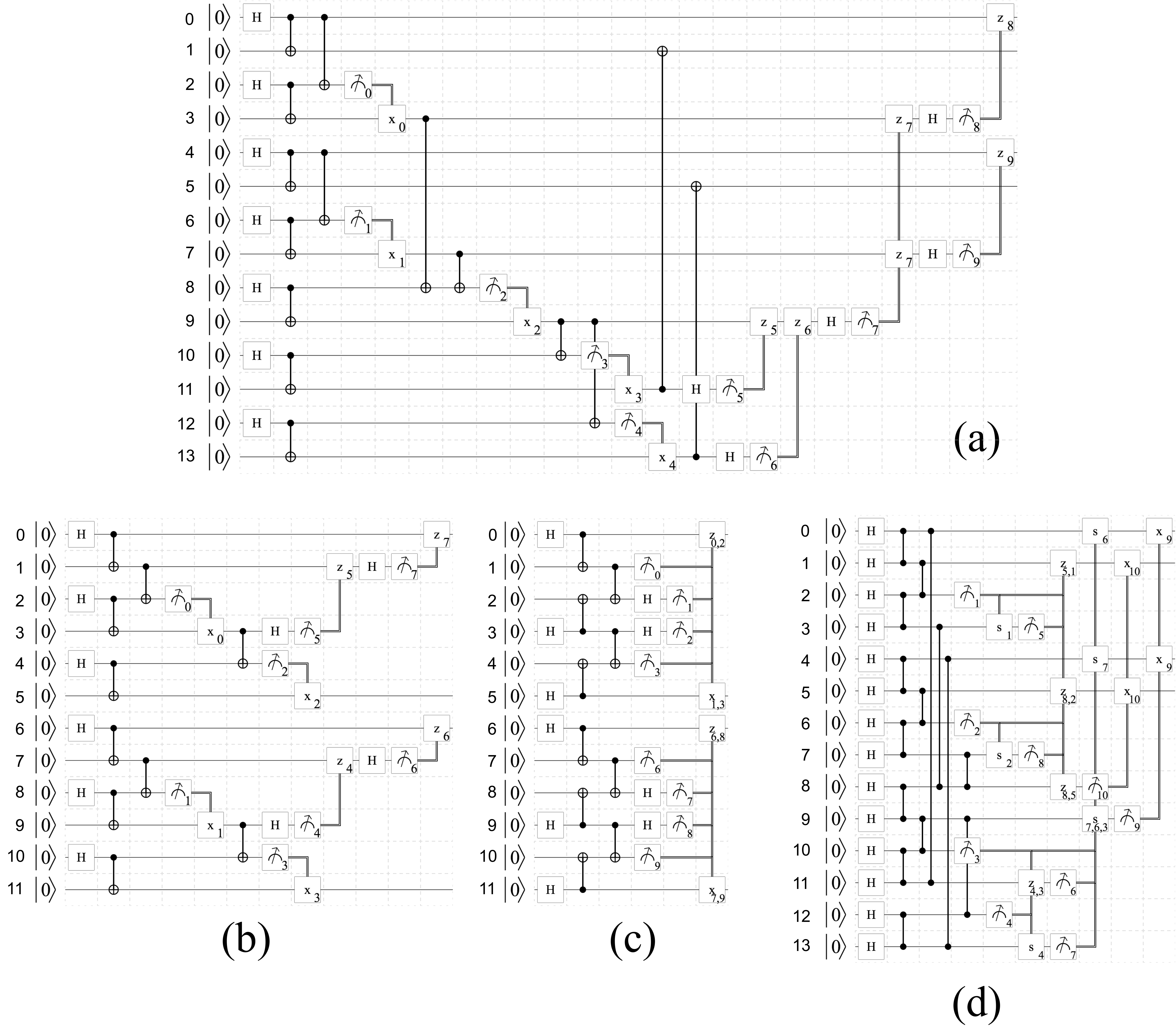}
  \caption{Evaluated quantum circuits. The first two steps of each circuit are the initialization part, which is independent from the protocol. (a) Quantum circuit for quantum network coding over repeater networks (QNC). (b) Quantum circuit for buffer space multiplexing using entanglement swapping (ES). (c) Quantum circuit for buffer space multiplexing using simultaneous entanglement swapping ($\mathrm{ES_{p}}$). (d) Quantum circuit for measurement-based quantum network coding over repeater networks (MQNC).}
  \label{circuits}
\end{figure*}

\begin{figure*}[!hbt]
  \center
  \includegraphics[keepaspectratio,scale=0.4]{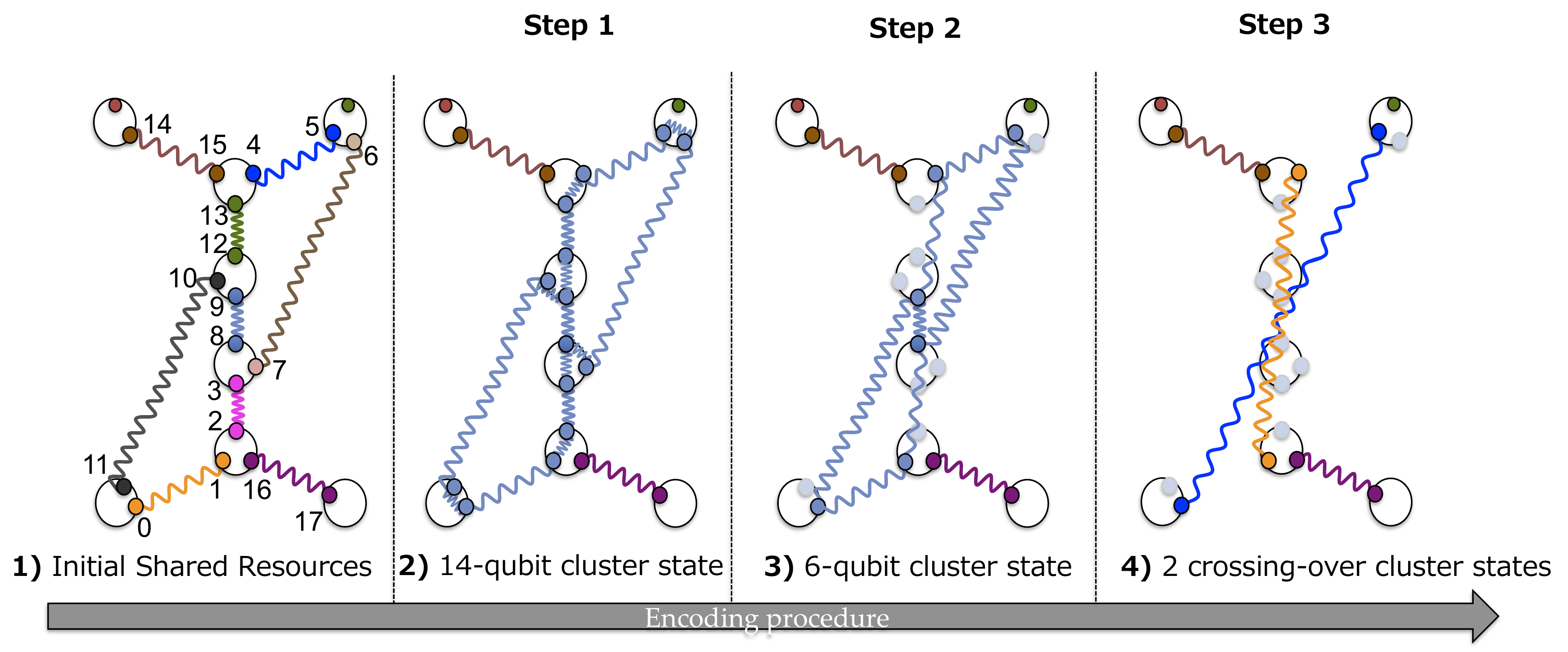}
  \caption{Measurement-based quantum network coding over grail network. For quantum channels across repeaters, the subgraph of the grail network equates to the butterfly network.}
  \label{grail}
\end{figure*}

This scheme can also be directly applied for the bottleneck problem that occurs in the quantum repeater grail network (see Fig. \ref{grail}), which is also one of the fundamental topologies with a bottleneck solvable via network coding \cite{Wang2007BeyondTB}.

\section{ANALYSIS}

We studied the behavior of the proposed protocol through Monte-Carlo simulation by tracking error propagations classically with various error sources. The results of the simulations are compared statistically with other alternative implementation methods, which include quantum network coding for repeater networks (QNC) and two types of buffer space multiplexing using entanglement swapping, one that performs entanglement swapping step-by-step based the classical message flow for path selection (ES) and another that performs all entanglement swapping operations simultaneously ($\mathrm{ES_{p}}$). The buffer space multiplexing assumes two links available at the bottleneck, therefore, both transmissions can be completed simultaneously using entanglement swapping (see Fig. \ref{buffer-space} for the pictorial model of ES). The circuit for ES is shown in Fig. \ref{circuits}(b) and the circuit for $\mathrm{ES_{p}}$ is in Fig. \ref{circuits}(c). In MQNC and $\mathrm{ES_{p}}$, we deferred byproduct operations to achieve fewer gates. The other circuits, QNC and ES, will be kept untouched and reused from the paper \cite{Satoh2016}.
Each protocol's statistical characteristics are summarized in Table \ref{stat}.

\begin{table}
\caption{Basic characteristics of protocols. The scaler KQ is calculated as the product of the number of qubits (Q) and the circuit depth (K) \cite{2003PhRvA..68d2322S}.}
\label{stat}
\begin{ruledtabular}
\begin{tabular}{ccccc}
& MQNC & QNC & ES & $\mathrm{ES_{p}}$ \\
\multirow{1}{*}{Number of qubits} & \multirow{1}{*}{14} & \multirow{1}{*}{14} & \multirow{1}{*}{12} & \multirow{1}{*}{12} \\
\multirow{1}{*}{Number of entangling operations} & \multirow{1}{*}{7} & \multirow{1}{*}{7} & \multirow{1}{*}{6} & \multirow{1}{*}{6} \\
\multirow{1}{*}{Number of single-qubit gates} & \multirow{2}{*}{14(14)} & \multirow{2}{*}{16(11)} & \multirow{2}{*}{12(8)} & \multirow{2}{*}{8(4)} \\
\multirow{1}{*}{(Byproduct operators)} & & & \\
\multirow{1}{*}{Number of two-qubit gates} & \multirow{1}{*}{8} & \multirow{1}{*}{8} & \multirow{1}{*}{4} & \multirow{1}{*}{4}  \\
\multirow{1}{*}{Number of measurements} & \multirow{1}{*}{10} & \multirow{1}{*}{10} & \multirow{1}{*}{4} & \multirow{1}{*}{4} \\
\multirow{1}{*}{Circuit depth} & \multirow{1}{*}{10} & \multirow{1}{*}{23} & \multirow{1}{*}{12} & \multirow{1}{*}{6} \\
\multirow{1}{*}{KQ} & \multirow{1}{*}{140} & \multirow{1}{*}{322} & \multirow{1}{*}{144} & \multirow{1}{*}{72} \\
\end{tabular}
\end{ruledtabular}
\end{table}

The first two error sources are the gate errors, which include the single-qubit gate error and the controlled gate error.
The third error source is faulty qubit measurement. An error on the measured qubit may also cause a faulty measurement result, which propagates to other qubits through misleading byproduct operations.
The fourth error is the memory error, which simulates the decoherence on qubits.
These errors will be applied to qubits per time step over the circuit.
The last error source is the initial resource error, which determines whether the state of a pre-shared entangled resource across two quantum repeaters, such as a Bell pair, is defective or not.

Only one operation per qubit is allowed in each time step. Consequently, for two different controlled gates with two different target qubits, if they share the same control qubit, a total of depth 2 at minimum is required to finish both operations.
Also, the physical distance between each node is not taken into consideration in the simulation. Thus, each node is assumed to be capable of perfectly delivering the classical feedforward message to the destination node within one time step.

In this paper, for simplicity, the term \emph{input fidelity} refers to the average fidelity of all pre-shared seven entangled pairs across repeaters and \emph{output fidelity} refers to the joint fidelity of the resulting two end-to-end entangled pairs at the end of each protocol.

Here, fidelity $F = 1 - p = \expval{\rho}{\psi}$, where $p$ is the error rate and $\ket{\psi}$ is the ideal pure state.
The output fidelity is calculated by $F_{output} = 1 - p'$, where $p'$ is the probability of at least one error being present on either entangled output at the end of the protocol.
For each datapoint, a maximum of 20 thousand residual errors have been accumulated or 1 million trials have been performed.

The rest of this section is constructed as follows. In subsection A, we first discuss the impact of the biased input error model with ideal local gate operations in all four protocols, one case with only $Z$ errors and another with only $X$ errors. Then, we discuss a more realistic model including all types of errors, such as $IY$ and $ZX$ error on Bell pair, keeping local gate operations ideal. In subsection B, we analyze the behavior of all four protocols under the total error model with imperfect initial resources and local gate operations.

\subsection{Input error propagation}

\begin{figure}[!hbt]
  \center
  \includegraphics[keepaspectratio,scale=0.35]{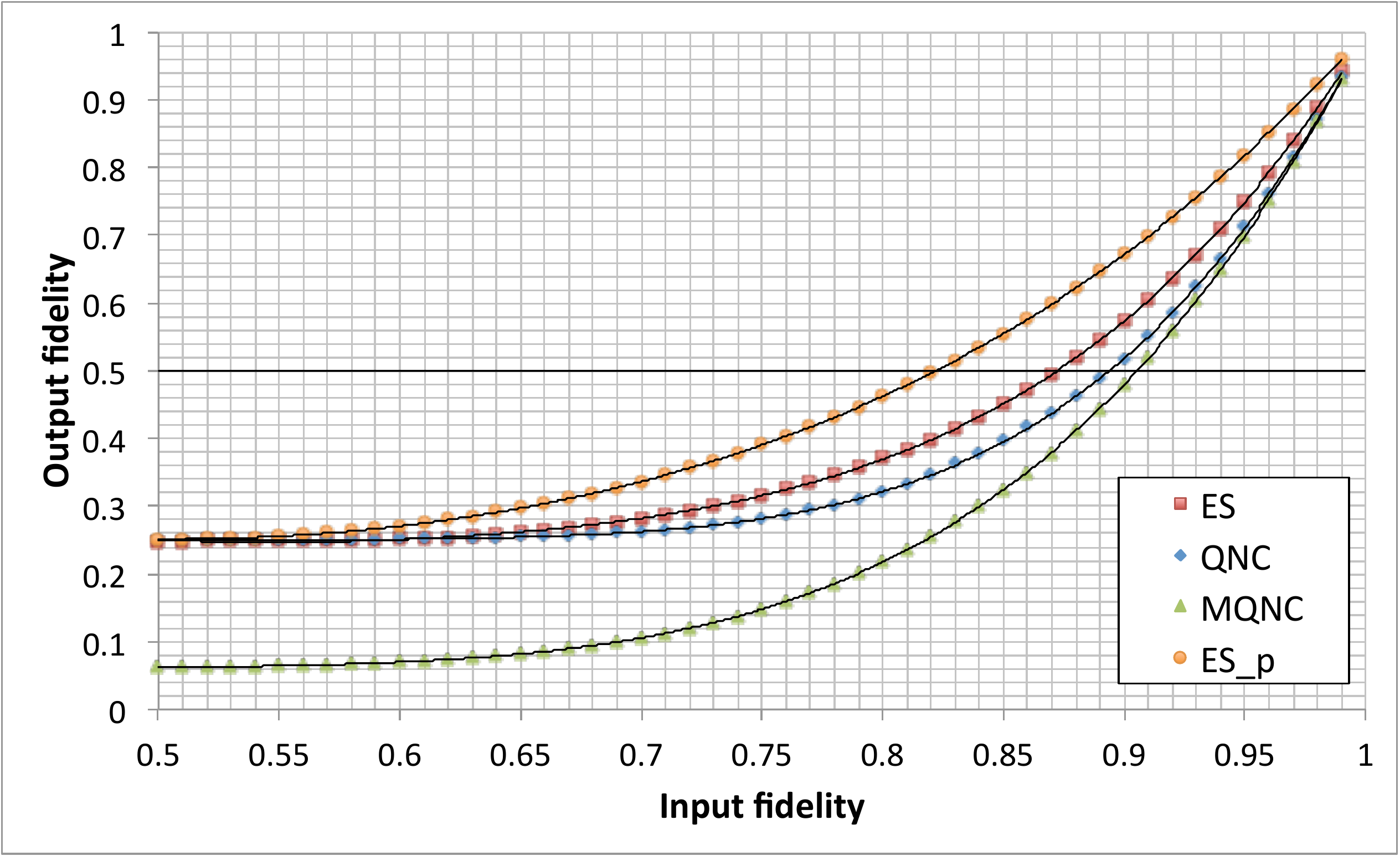}
  \caption{Impact of input fidelity on output fidelity in four protocols. Probabilistic $Z$ error on qubits labeled with odd numbers. Local operations are assumed to be ideal.}
  \label{mqnc_initZ}
\end{figure}

The first scenario simulates the artificial input error model with only $Z$ errors stochastically present on the pre-shared entangled pairs of qubits. Local gate operations are assumed to be ideal. For simplicity, errors are assumed to only exist on the qubits that are labeled with odd numbers, after the initialization, as in Fig. \ref{circuits}.
As an example, the initial entangled resource of qubit 0 and qubit 1 may have a state of $ (I_{0} \otimes Z_{1}) \ket{\psi_{0,1}} $ with probability $ P_{error} = p $, or $ (I_{0} \otimes I_{1}) \ket{\psi_{0,1}} $ with probability $P_{clean} = 1 - p$.
The simulation result is shown below at Fig. \ref{mqnc_initZ}.

Overall, $\mathrm{ES_{p}}$ has the highest initial resource $Z$ error tolerance among the protocols.
ES and QNC have similar results with lower and higher fidelity, and the difference between ES and QNC output fidelity becomes $\mid F_{output}^{qnc} - F_{output}^{es} \mid \geq 1\%$ when $67\% \leq F_{input} \leq 98\%$.
QNC, ES and $\mathrm{ES_{p}}$ have higher output fidelity compared to MQNC because with this biased input error model, for each output pair, one-fourth of the input error combinations result in stabilizing the output state.
On the other hand, no $Z$ errors on initial resources end up stabilizing the cluster state in MQNC.

\begin{figure}[!hbt]
  \center
  \includegraphics[keepaspectratio,scale=0.35]{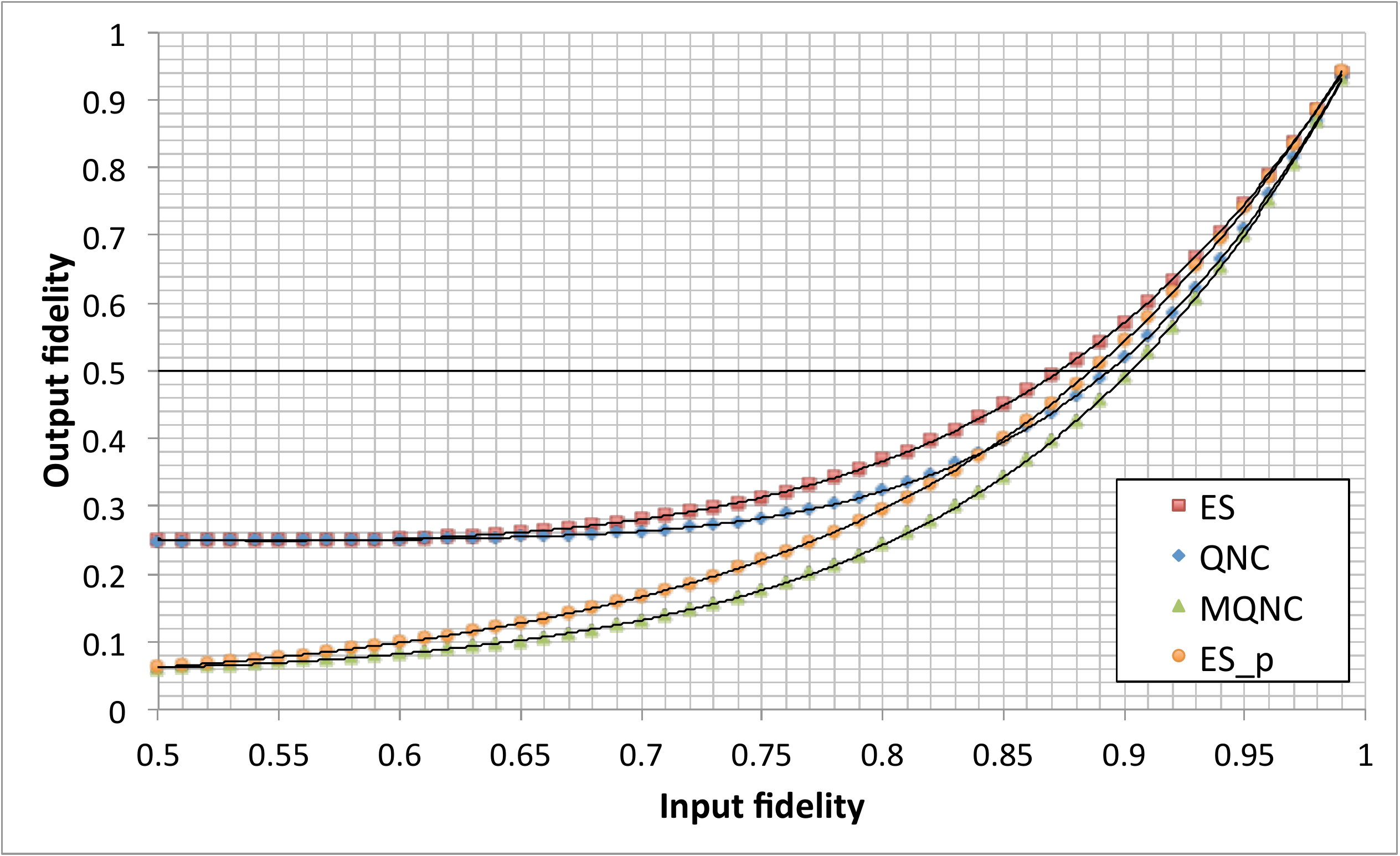}
  \caption{Impact of input fidelity on output fidelity in four protocols. Probabilistic $X$ error on qubits labeled with odd numbers. Local operations are assumed to be ideal.}
  \label{mqnc_initX}
\end{figure}

The second scenario is similar to the first scenario but with $X$ errors present on the pre-shared entangled pairs of qubits labeled with odd numbers. The simulation result is shown in Fig. \ref{mqnc_initX}.

The $X$ error tolerance of QNC and ES is symmetric to the case with only $Z$ errors. In contrast, MQNC has slightly better $X$ error tolerance than $Z$ error. The $X$ error tolerance of $\mathrm{ES_{p}}$ drops significantly from the case with only $Z$ error.
For reasons similar to the case with only $Z$ error, this biased error model results in a joint fidelity of $1/16=6.25\%$ for MQNC and $\mathrm{ES_{p}}$ and $1/4=25\%$ for QNC and ES, when input fidelity is minimum $F_{input}=50\%$. The difference in MQNC and QNC output fidelity $\mid F_{output}^{mqnc} - F_{output}^{qnc} \mid \leq 1\%$ when $F_{input} \geq 96\%$.

If a physical system has a biased error model, ES might be stronger than other three protocols.

Finally, not only $IZ$ or $IX$ errors but any other errors, such as $ZZ$ error on a 2-qubit cluster state, can exist on any initial resources. Errors on entangled resources have the same weighted probability. This scenario tells us the overall input error sensitivity for each protocol.

\begin{figure}[!hbt]
  \center
  \includegraphics[keepaspectratio,scale=0.35]{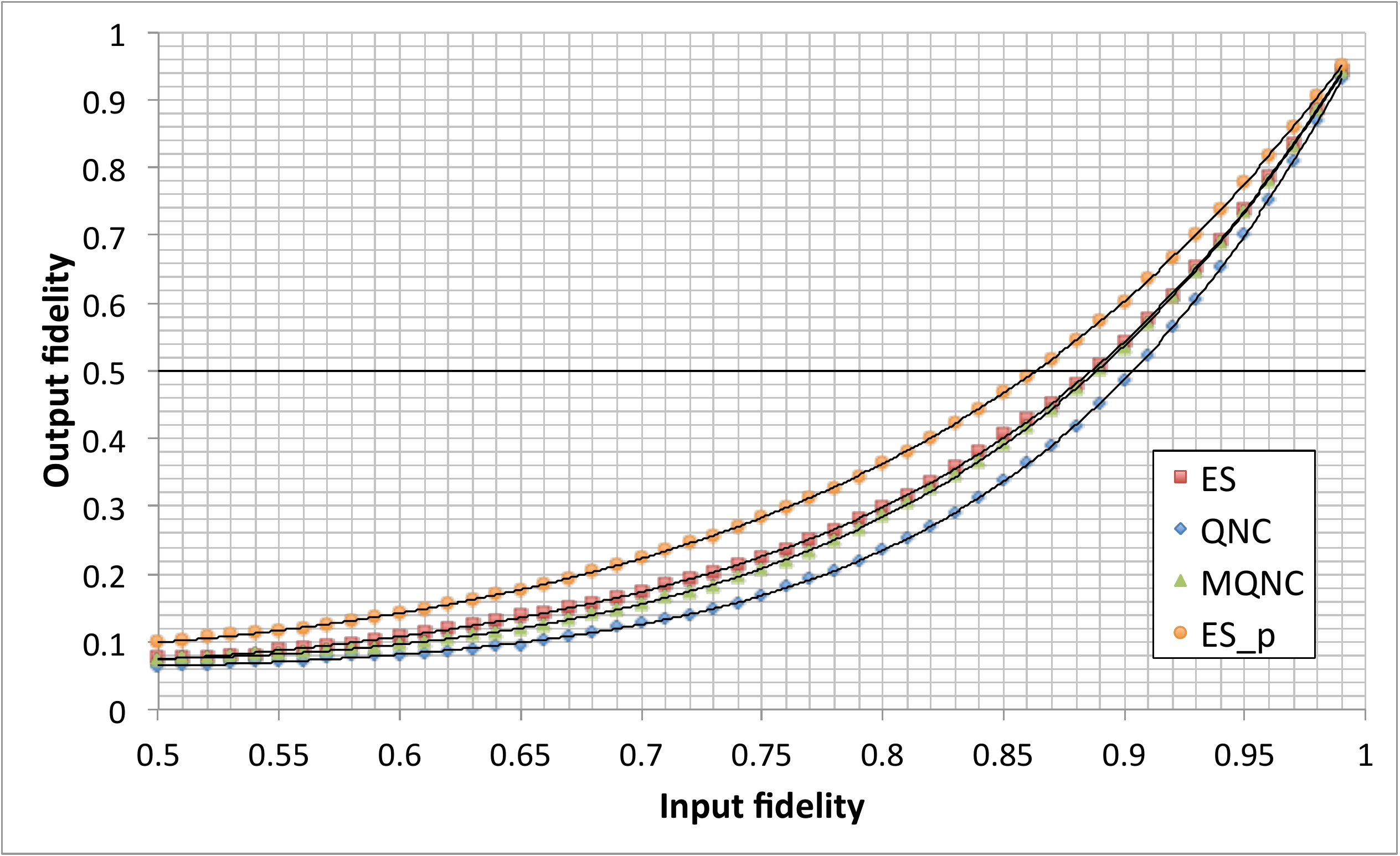}
  \caption{Impact of input fidelity on output fidelity in four protocols. All combinations of observable errors stochastically present on all qubits. Local operations are assumed to be ideal.}
  \label{mqnc_initErr}
\end{figure}

As shown in Fig. \ref{mqnc_initErr}, MQNC and ES have similar initial error tolerance, and QNC is slightly left behind while $\mathrm{ES_{p}}$ is slightly ahead. In order to retain an output fidelity of $F_{output} = 50\% $, $\mathrm{ES_{p}}$ requires an input fidelity of at least $F_{input} =87\%$, ES and MQNC require at least $F_{input} =89\%$, while QNC requires an extra 2\% for achieving the same goal.

Most input errors in QNC, ES and $\mathrm{ES_{p}}$ develop as $X$ error, $Z$ error or $XZ$ error. The error distribution caused by input errors in MQNC is equally weighted to all types (for details refer to Appendix \ref{input_error}).

\subsection{Total error model}

\begin{figure}[!htb]
  \center
  \includegraphics[keepaspectratio,scale=0.35]{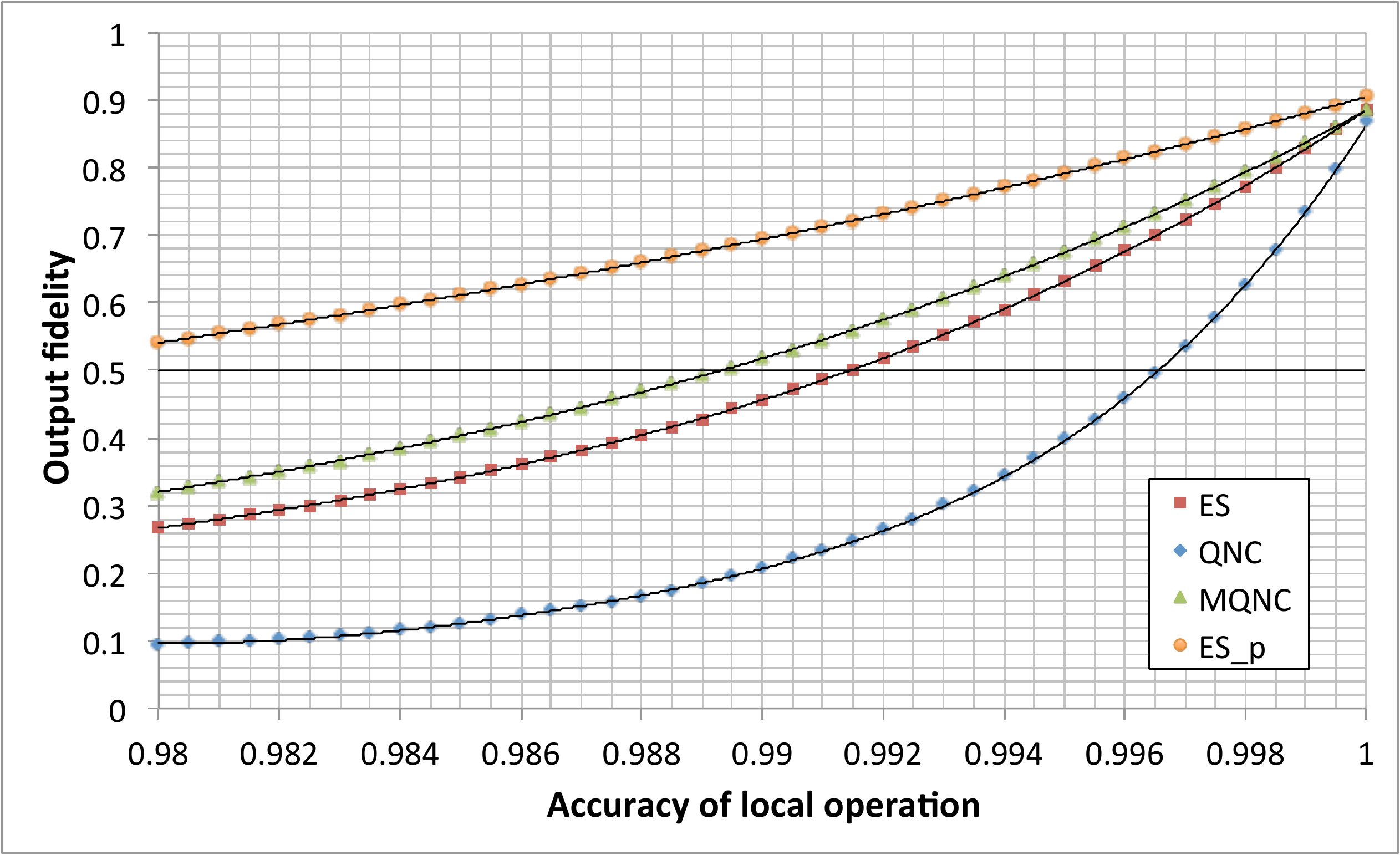}
  \caption{Impact of local operation accuracy on output fidelity in four protocols. Input fidelity is fixed to $F_{input} = 98\%$, and the local operation accuracy is changed from $F_{operation} = 98\%$ to $F_{operation} = 100\%$ with $\Delta F_{operation} = 0.05\%$.}
  \label{98_allErr}
\end{figure}

\begin{figure*}[!htb]
  \center
  \includegraphics[width=0.6\linewidth]{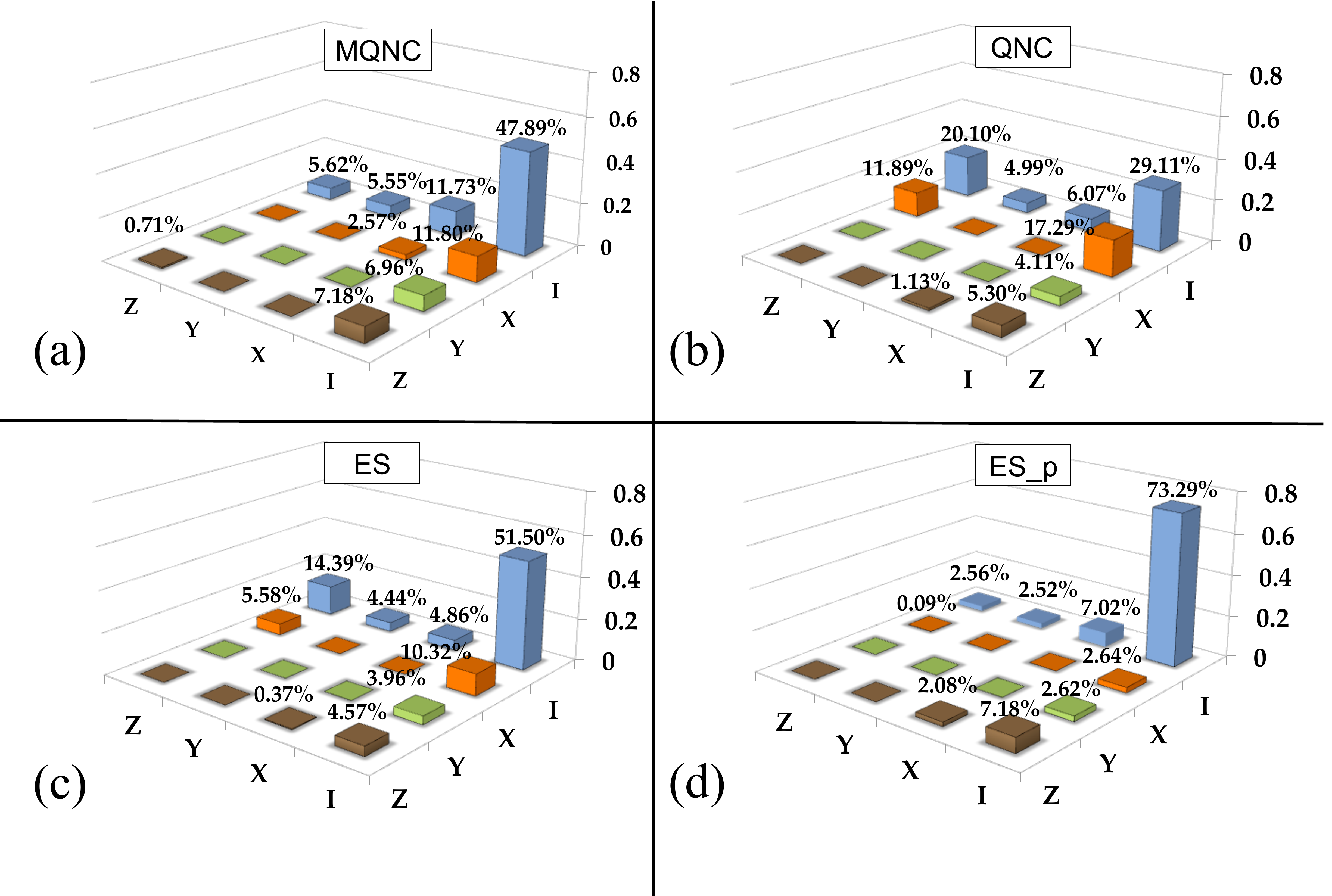}
  \caption[Error distribution on entangled outputs]{Error distribution on entangled outputs. (a) Error distribution of end-to-end cluster state composed of qubit 0 and qubit 5 in MQNC. (b) Error distribution of end-to-end Bell pair composed of  qubit 0 and qubit 5 in QNC.(c) Error distribution of end-to-end Bell pair composed of qubit 0 and qubit 5 in ES. (d) Error distribution of end-to-end Bell pair composed of qubit 0 and qubit 5 in n $\mathrm{ES_{p}}$. Distribution at input fidelity $F_{input} = 98\%$ and local operation accuracy $F_{operation} = 98\%$. The error distribution is symmetrical over 2 outputs for all protocols.}
  \label{error_dist_9898}
\end{figure*}

Finally, we consider the comprehensive error model, with all sources and error types included. A single-qubit operation may emit one error out of three possibilities, $X$, $Y$ and $Z$ error, with equal probability. Similarly, after a two-qubit operation, there is no
error, or at least one $X$, $Y$ or $Z$ error is present on either qubit. Simulations have shown that purifying Bell pairs to $F = 98\%$ is feasible \cite{Meter2009SystemDF}.
Therefore fidelity for pre-shared resources are assumed to be maintained at $F_{input} = 98\%$ during waiting time.
Other local operation error rates are changed concurrently with equal magnitude from $F_{operation} = 98\%$ to $F_{operation} = 100\%$ with $\Delta F_{operation} = 0.05\%$. The simulation result is shown in Fig. \ref{98_allErr} and Fig. \ref{error_dist_9898}.

As shown, with the model of all error sources, $\mathrm{ES_{p}}$ obtains the highest output fidelity at any local operation accuracy, and MQNC is in between $\mathrm{ES_{p}}$ and ES.
Although MQNC, ES and $\mathrm{ES_{p}}$ tolerate more than twice the local error rate of QNC, the output fidelity of each protocol reaches to a similar point with a sufficiently high local operation accuracy.
The output infidelity when $F_{operation}=99.95\%$ is 13.9\% for MQNC, 10.7\% for $\mathrm{ES_{p}}$, 14.4\% for ES and 20.2\% for QNC. The output fidelity of QNC is more sensitive to the local error rate than the other three protocols.

\begin{figure}[!htb]
  \center
  \includegraphics[keepaspectratio,scale=0.35]{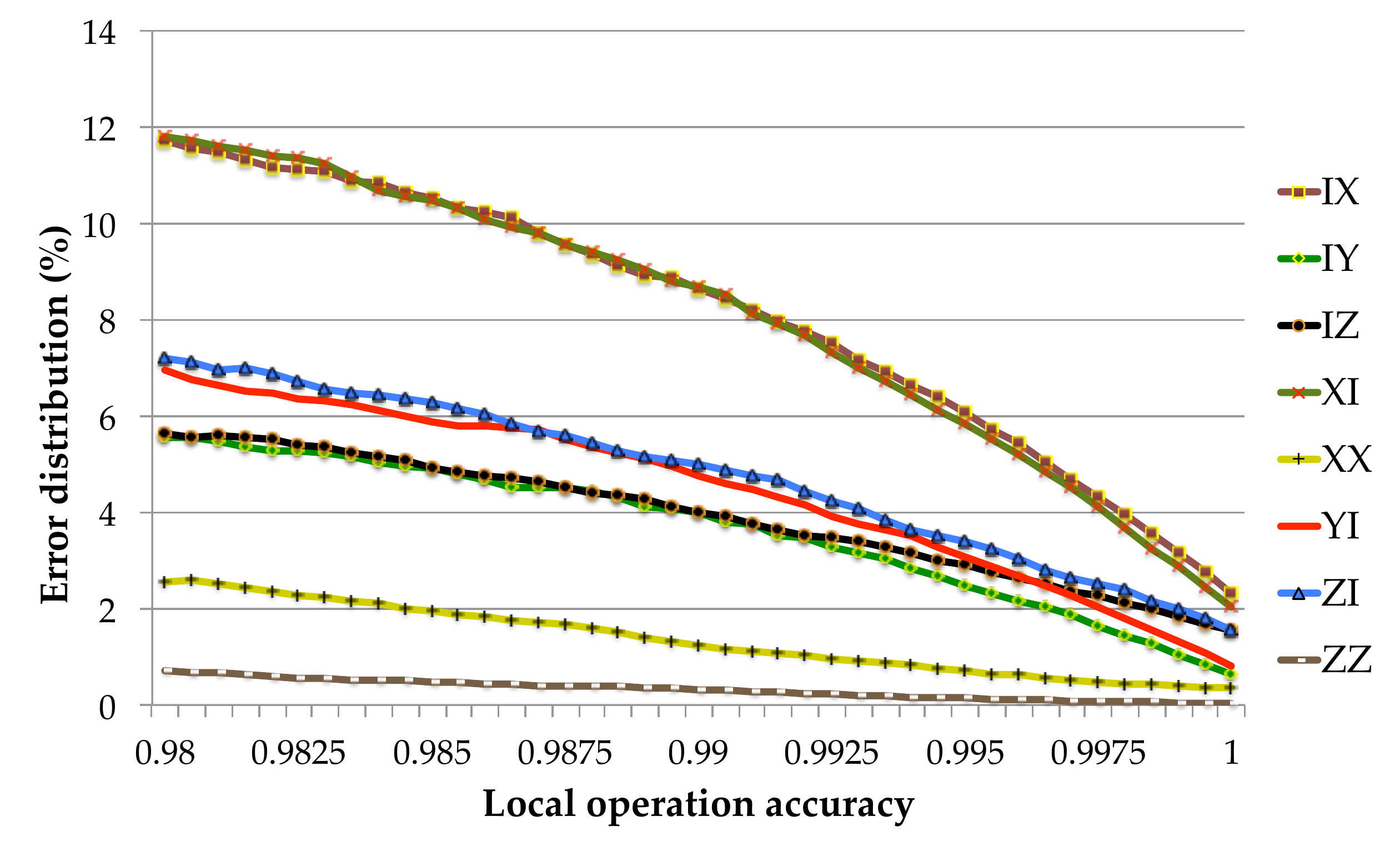}
  \caption{Error distribution versus local operation accuracy of output cluster state composed of qubit 0 and qubit 5 in MQNC. Input fidelity is fixed to $F_{input} = 98\%.$}
  \label{error_dist_change_mqnc}
\end{figure}

The change in error distribution over the change in local operation accuracy for MQNC is plotted in Fig. \ref{error_dist_change_mqnc}.

As the result shows, MQNC mostly suffers from $IX$ and $XI$ error in all situations - a bit-flip error on either qubit. Those errors combined account for approximately one fourth of the total probability.
On the other side, with a fixed initial resource error of $F_{input}=98\%$, the probability of $ZZ$ and $XX$ error drops gradually to $P(ZZ) \approx 0.03\%$ and $P(XX) \approx 0.36\%$ as local operation accuracy approaches to 1. As those error rates converge to a certain point, not much benefit can be obtained from further improvement of local operation accuracy, when the local operation accuracy is high enough. The error rate for $ZZ$ error drops by $0.006\%$ by an improvement of local operation accuracy $F_{operation}=99.995\%$ to $F_{operation}=100\%$. While the slope of $ZZ$ error and $XX$ error gets flatter, other error types' probabilities drop more aggressively as the local operation accuracy approaches to 1. The $IX$ error rate decreases by $0.46\%$ when local operation accuracy is improved from $F_{operation}=99.995\%$ to $F_{operation}=100\%$.

\subsection{Total error model with ideal qubit memories}
\begin{figure}[!hbt]
  \center
  \includegraphics[keepaspectratio,scale=0.35]{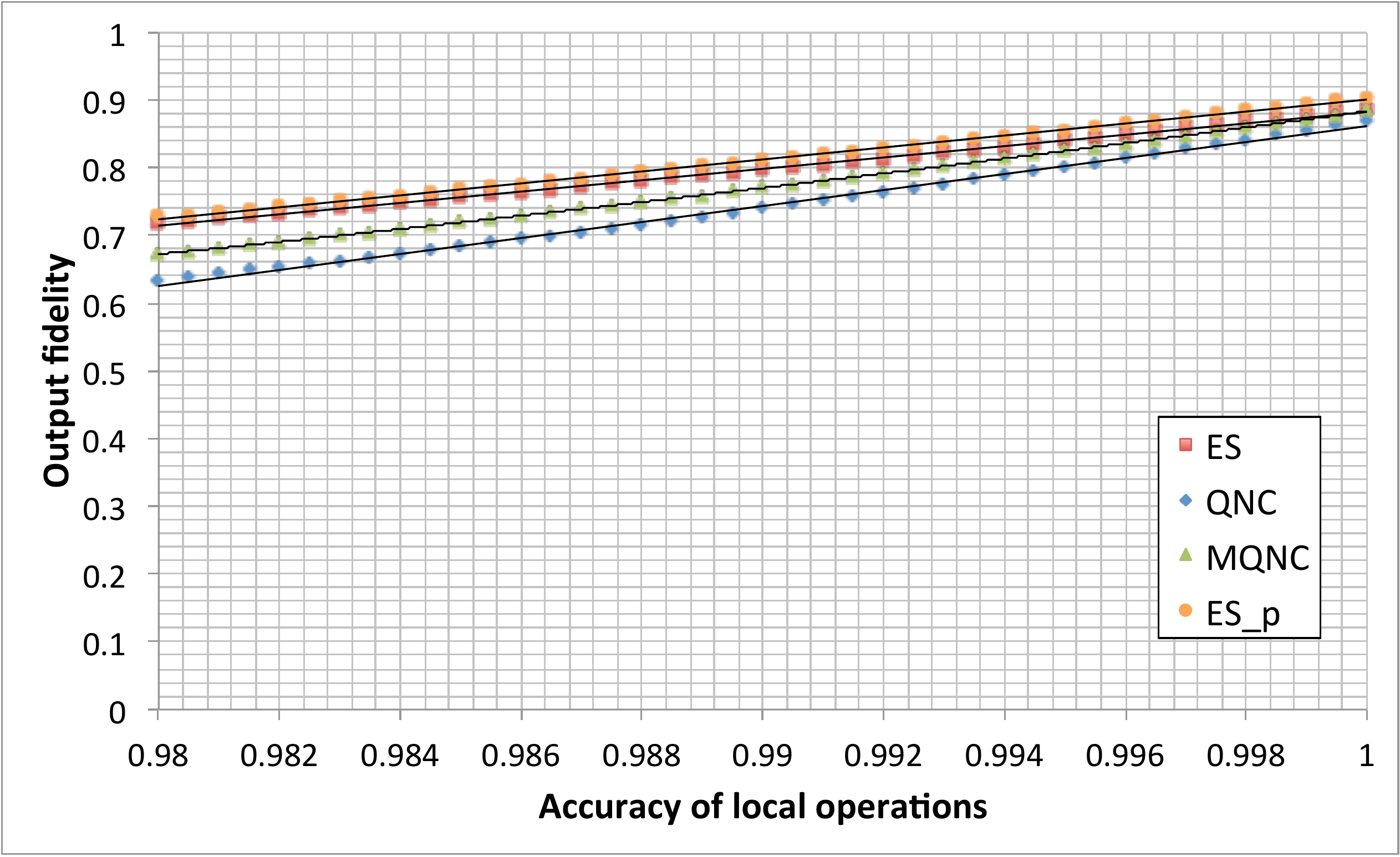}
  \caption{Impact of local operation accuracy on output fidelity in four protocols.  Memory error rate is fixed to $F_{memory} = 100\%$. Input fidelity is fixed to $F_{input} = 98\%$.}
  \label{no_memory_error}
\end{figure}

Finally, it is worth separating the effect of gate errors from memory errors.
Qubit memories are assumed to be ideal, and therefore, idle qubits are immune to noise. Other error variables including gates and measurements stay as the independent variable with a domain of $F_{operation} = 98\%$ to $F_{operation} = 100\%$ and a constant initial resource fidelity of $F_{input} = 98\%$. The simulation results are shown in Fig. \ref{no_memory_error}.

Unlike the simulation results with the total error model, both protocols based on entanglement swapping obtain a higher output fidelity than network coding. All four protocols end up with similar output fidelities as the local operation accuracy approaches to 1. While $\mathrm{ES_{p}}$ obtains the highest error tolerance with better qubit memories, in all four protocols, memory imperfection is the dominant error and is the main causes of faulty communication.

\begin{figure}[!hbt]
  \center
  \includegraphics[keepaspectratio,scale=0.35]{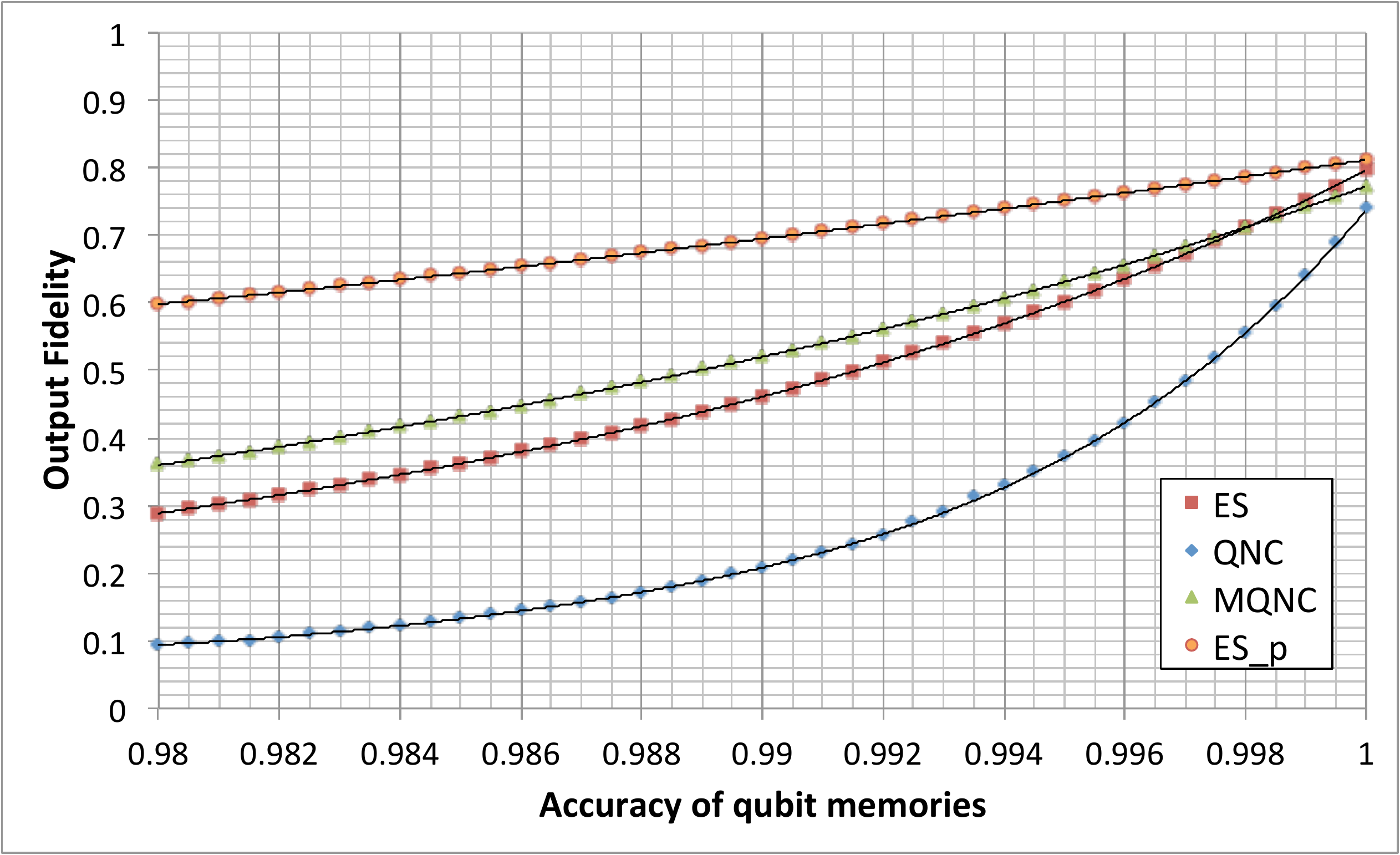}
  \caption{Impact of qubit memory imperfection on output fidelity in four protocols.  Local operation accuracy, excluding memory, is fixed to $F_{operation} = 99\%$. Input fidelity is fixed to $F_{input} = 98\%$.}
  \label{memChange}
\end{figure}

Lastly, gate and measurement accuracies are fixed to $F_{operation} = 99\%$, initial resource fidelity is fixed to $F_{initial} = 98\%$, and memory accuracy is changed from $F_{memory} = 98\%$ to $F_{memory} = 100\%$ using $\Delta F_{memory} = 0.05\%$, in order to assess the impact of memory accuracy on the protocol robustness. The simulation result is shown in Fig. \ref{memChange}.

Because the robustness of MBQC mainly comes from shorter circuit depth, the higher the memory error rate relative to other error sources, the bigger the gap in output fidelity. The output fidelity of MQNC is higher than that of ES when $F_{memory} \leq 99.8\%$.

\section{CONCLUSION}

Using Monte-Carlo simulation, this paper discussed the simulated error propagation of four different protocols, MQNC, QNC, ES and $\mathrm{ES_{p}}$ on a butterfly network.
This work has not attempted to prove optimality, and therefore results in this paper are specific to the protocols in Fig. \ref{circuits}.
Thus, the issue of optimal circuit design for both classes, MQNC and QNC, still remains an open question.

MQNC is more sensitive to $Z$ errors, and has no practical advantage over ES in terms of initial error tolerance with ideal local gate operations.
Unlike QNC and ES, the correlation between the input and output fidelity differs from $Z$ errors to $X$ errors, as only $X$ errors propagate through CZ gates. In the asymptotic limit with the artificial model of only a single error type, MQNC fares worse than either QNC, ES and $\mathrm{ES_{p}}$ because both $X$ and $Z$ errors develop in the final 2-qubit cluster states.
In general, however, our measurement-based quantum repeater network coding scheme significantly simplified the network coding procedure and showed a substantial improvement of overall error tolerance compared to QNC, and is even slightly better than ES with the total error model.

One should also be reminded that buffer space multiplexing requires an extra entangled state at the bottleneck to complete both transmissions simultaneously.
As a conclusion, MQNC is more broadly applicable than QNC, but the choice of MQNC or buffer space multiplexing still depends on the situation and the network topology. If resources for networking are abundant, ES may be more useful. In contrast, MQNC is more practical for resolving critical resource contentions over networks.

\begin{acknowledgments}
This work is supported by JSPS KAKENHI Grant Number 16H02812.
\end{acknowledgments}

\bibliography{Analysis_of_Measurement-based_Quantum_Network_Coding_over_Repeater_Networks_under_Noisy_Conditions}

\appendix

\section{Step-by-step explanation of quantum network coding for repeater networks}
\label{qnc_step}

A six-step procedure takes us from the seven separate Bell pairs to two end-to-end Bell pairs via QNC.

We begin with

\begin{equation}
  \ket{\Psi_{0}} = \ket{\Phi^{+}_{0,1}}\ket{\Phi^{+}_{2,3}}\ket{\Phi^{+}_{4,5}}\ket{\Phi^{+}_{6,7}}\ket{\Phi^{+}_{8,9}}\ket{\Phi^{+}_{10,11}}\ket{\Phi^{+}_{12,13}},
\end{equation}

where the subscripts correspond to the numbered qubits in Fig. \ref{qnc}.

Using the given resources, the first step of the protocol connects a particular set of Bell pairs to generate two 3-qubit GHZ states. Therefore, the overall system after Step 1 can be described as:

\begin{equation}
\ket{\Psi_{1}} = \ket{GHZ_{0,1,3}} \ket{GHZ_{4,5,7}} \ket{\Phi^{+}_{8,9}} \ket{\Phi^{+}_{10,11}} \ket{\Phi^{+}_{12,13}}.
\end{equation}

The bottleneck link is manipulated to bridge the GHZ states to the left and right via a parity measurement as in Step 2:

\begin{eqnarray}
{\ket{\Psi_{2}}}=&& \ket{\Phi^{+}_{10,11}} \ket{\Phi^{+}_{12,13}} \otimes  \nonumber \\
&& \left( \frac{1}{2}(\ket{0_{0}0_{1}0_{3}0_{4}0_{5}0_{7}} + \ket{1_{0}1_{1}1_{3}1_{4}1_{5}1_{7}})\ket{0_{9}} \right. \nonumber \\
&& \left. + \frac{1}{2}(\ket{0_{0}0_{1}0_{3}1_{4}1_{5}1_{7}} + \ket{0_{0}0_{1}0_{3}1_{4}1_{5}1_{7}})\ket{1_{9}} \right). \nonumber \\
\end{eqnarray}

Step 3 requires the FANOUT operations, which generally involves an arbitrary quantum state and two Bell pairs such as:
\begin{equation}
\ket{\psi_{before}} = (\alpha \ket{0_{0}} + \beta \ket{1_{0}})\ket{\Phi^+_{1,2}}\ket{\Phi^+_{3,4}}.
\end{equation}

Using the given resources, the FANOUT operation results in a system:
\begin{equation}
\ket{\psi_{after}} = (\alpha \ket{000_{0,2,4}} + \beta \ket{111_{0,1,3}}).
\end{equation}

In Step 3, the FANOUT operation is applied to the parity qubit, which in this case is qubit 9:

\begin{eqnarray}
{\ket{\Psi_{3}}}=&& \frac{1}{2}(\ket{0_{0}0_{1}0_{3}0_{4}0_{5}0_{7}} + \ket{1_{0}1_{1}1_{3}1_{4}1_{5}1_{7}})\ket{0_{9}0_{11}0_{13}}  \nonumber \\
&& + \frac{1}{2}(\ket{0_{0}0_{1}0_{3}1_{4}1_{5}1_{7}} + \ket{0_{0}0_{1}0_{3}1_{4}1_{5}1_{7}})\ket{1_{9}1_{11}1_{13}}. \nonumber \\
\end{eqnarray}

In Step 4, the CNOT operation will be applied to qubits at both target nodes:

\begin{eqnarray}
{\ket{\Psi_{4}}}=&& \frac{1}{2}(\ket{0_{0}0_{1}0_{3}0_{4}0_{5}0_{7}} + \ket{1_{0}1_{1}1_{3}1_{4}1_{5}1_{7}})\ket{0_{9}0_{11}0_{13}}  \nonumber \\
&& + \frac{1}{2}(\ket{0_{0}1_{1}0_{3}1_{4}0_{5}1_{7}} + \ket{1_{0}0_{1}1_{3}0_{4}1_{5}0_{7}})\ket{1_{9}1_{11}1_{13}}. \nonumber \\
\end{eqnarray}

In Step 5, parity qubits at both target nodes are removed from the graph by performing $Z$ measurements:

\begin{eqnarray}
{\ket{\Psi_{5}}}=&& \frac{1}{2}(\ket{0_{0}0_{1}0_{3}0_{4}0_{5}0_{7}} + \ket{1_{0}1_{1}1_{3}1_{4}1_{5}1_{7}})\ket{0_{9}}  \nonumber \\
&& + \frac{1}{2}(\ket{0_{0}1_{1}0_{3}1_{4}0_{5}1_{7}} + \ket{1_{0}0_{1}1_{3}0_{4}1_{5}0_{7}})\ket{1_{9}}. \nonumber \\
\end{eqnarray}

Similarly, the parity qubit at the bottleneck link is removed in Step 6:

\begin{eqnarray}
{\ket{\Psi_{6}}}=&& \frac{1}{2}(\ket{0_{0}0_{1}0_{3}0_{4}0_{5}0_{7}} + \ket{1_{0}1_{1}1_{3}1_{4}1_{5}1_{7}})  \nonumber \\
&& + \frac{1}{2}(\ket{0_{0}1_{1}0_{3}1_{4}0_{5}1_{7}} + \ket{1_{0}0_{1}1_{3}0_{4}1_{5}0_{7}}). \nonumber \\
\end{eqnarray}

In the last step, the remaining qubits at the bottleneck node are removed from the graph to form two cross-over Bell pairs:

\begin{eqnarray}
{\ket{\Psi_{7}}}=&& \frac{1}{2}(\ket{0_{0}0_{1}0_{4}0_{5}} + \ket{1_{0}1_{1}1_{4}1_{5}}) \nonumber \\
&& + \frac{1}{2}(\ket{0_{0}1_{1}1_{4}0_{5}} + \ket{1_{0}0_{1}0_{4}1_{5}}) \nonumber \\
&& = \ket{\Phi^{+}_{0,5}} \otimes \ket{\Phi^{+}_{1,4}}.
\end{eqnarray}

This completes the sequence and results in two end-to-end Bell pairs, which can be used to teleport the message qubit from source to destination directly.

\section{Propagation of Input errors}
\label{input_error}

Let us explain in more detail the propagation of errors from the
initial Bell pairs to the output states and the behavior of errors
that cancel out, acting as stabilizers on the output states.  For
simplicity, we review the behavior of all four protocols for the
conditions corresponding to the left edge of Fig. \ref{mqnc_initErr}: local operations are perfect ($F_{operation} = 100\%$), but for illustration the
fidelity of the initial two-qubit entangled states is only 50\%
($F_{input} = 50\%$).

\begin{figure*}[!hbt]
  \center
  \includegraphics[width=0.6\linewidth]{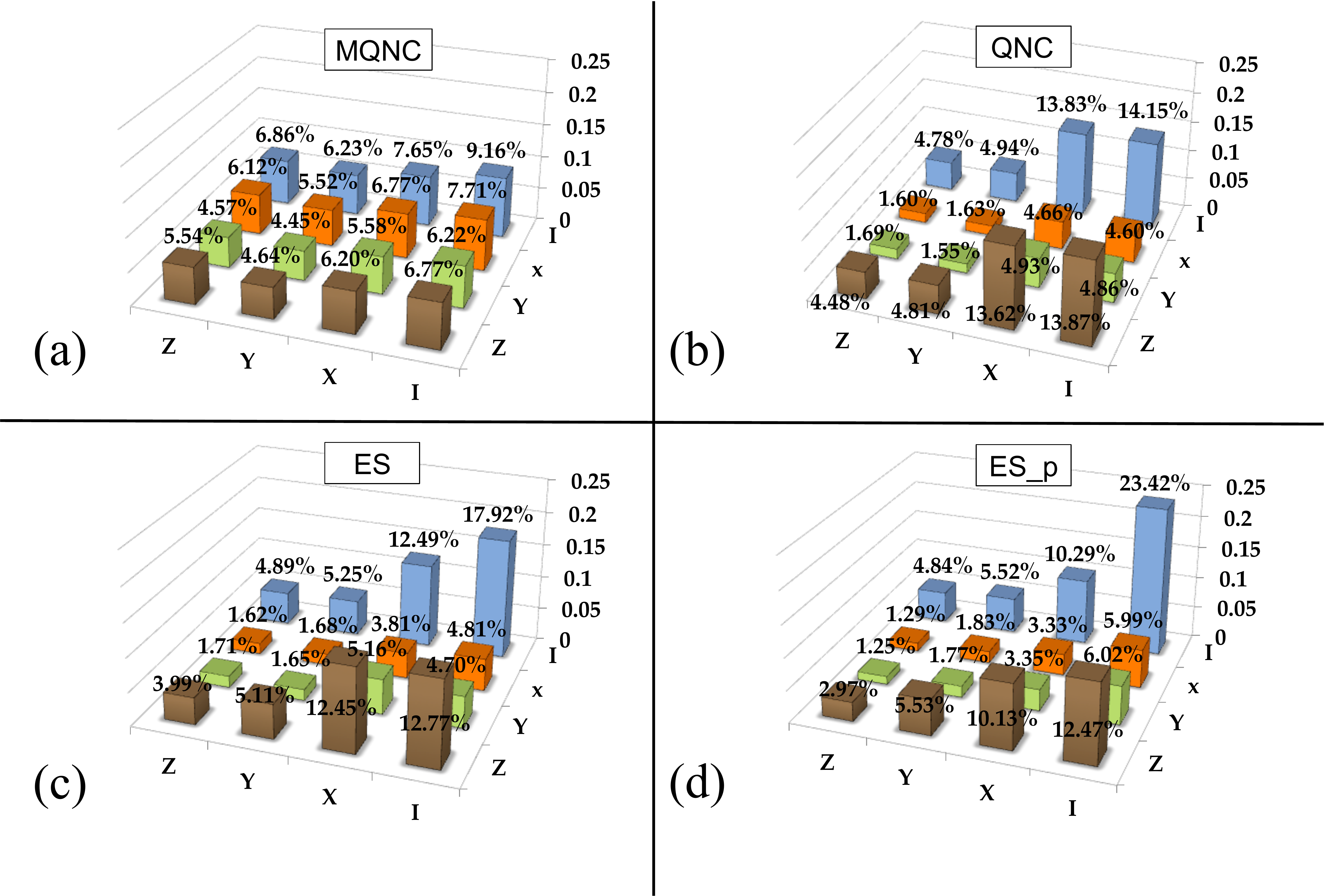}
  \caption{Accumulated error types developed throughout the circuit from input errors in all protocols. Local operation accuracy is $F_{operation} = 100\%$ and input fidelity is $F_{input} = 50\%$.}
  \label{input_raw}
\end{figure*}

\begin{figure*}[!hbt]
  \center
  \includegraphics[width=0.6\linewidth]{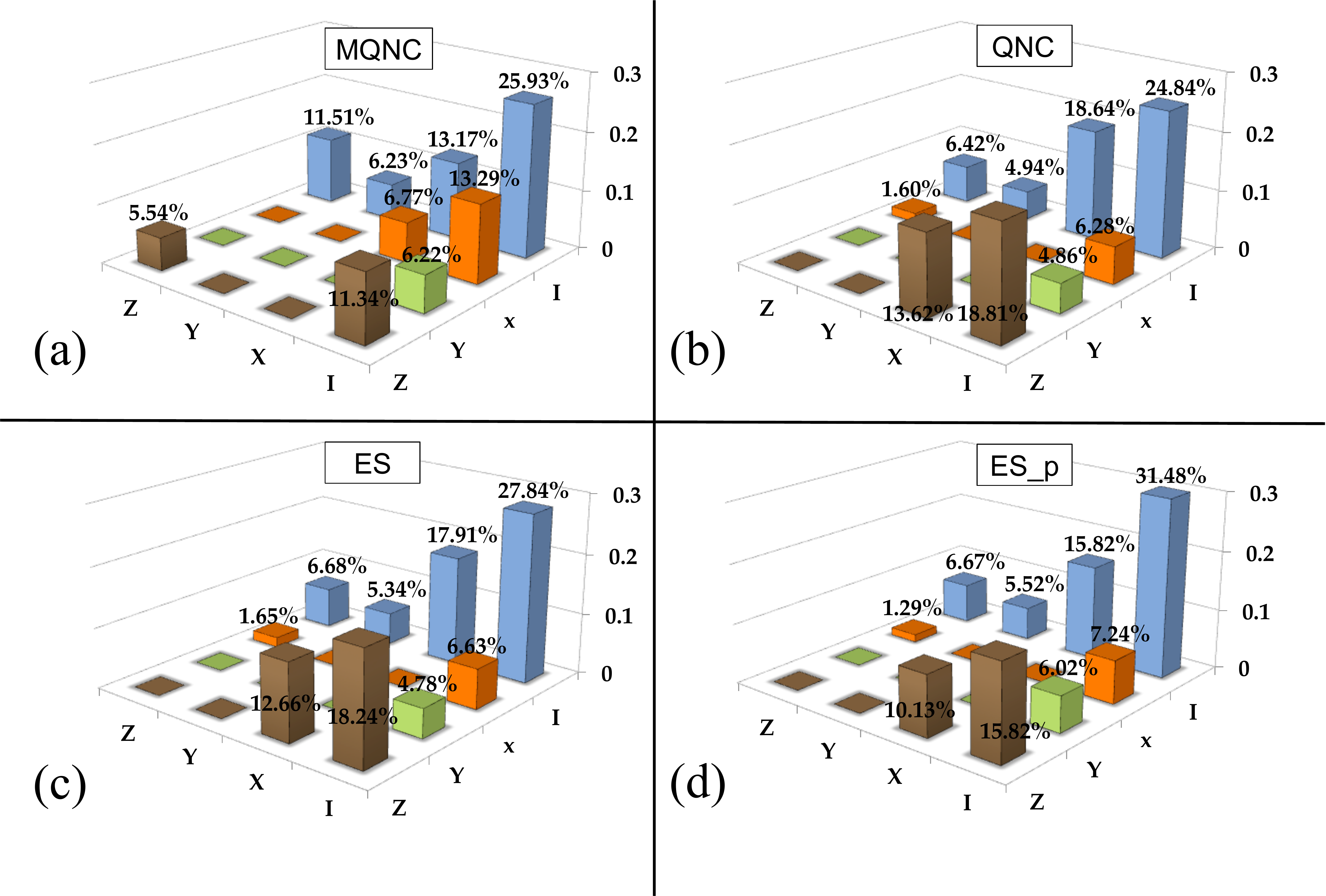}
  \caption{The error distribution of the one output state for all protocols. Local operation accuracy is $F_{operation} = 100\%$ and input fidelity is $F_{input} = 50\%$.}
  \label{input_stab}
\end{figure*}

Fig. \ref{input_raw} shows the direct action of errors introduced by the simulator, before taking into account that certain error patterns are
stabilizers.
In the entanglement swapping protocols, fewer initial Bell pairs influence each output Bell pair, so the output fidelity is slightly different from the measurement-based protocols.  In practice, the errors that are stabilizers (e.g., $ZZ$ errors for $\mathrm{ES_{p}}$) are indistinguishable from $II$, and in Fig. \ref{input_stab} are folded back into the $II$ state.  The exact error patterns differ, but in all four protocols the output states are stabilized two-qubit states, so the same number of error cases result in no detectable error.  If the probability of $X$, $Y$, and $Z$ errors are the same, therefore, the final fidelities are not very different, but with asymmetric error processes, the results can be rather different, as shown in Figs. \ref{mqnc_initZ}, \ref{mqnc_initX} and \ref{mqnc_initErr}.

\end{document}